\documentclass[conference]{IEEEtran}
\usepackage{listings}
\usepackage{color}
\usepackage[T1]{fontenc}
\usepackage{lmodern}
\usepackage{multicol}
\usepackage{mathabx}
\usepackage{multirow}
\usepackage[numbers]{natbib}
\newcommand{\bnote}[2]{
	\fbox{\bfseries\sffamily\scriptsize#1}
	{\textsc\small$\blacktriangleright$\textit{#2}$\blacktriangleleft$}}

\newif\ifcomment
\commenttrue
\ifcomment
\newcommand{\liliane}[1]{\bnote{Liliane}{\textcolor[rgb]{1,0,1}{#1}}}
\newcommand{\marcelo}[1]{\bnote{Marcelo}{\textcolor{cyan}{#1}}}
\else
\newcommand{\liliane}[1]{}
\newcommand{\marcelo}[1]{}
\fi

\ifCLASSINFOpdf
\else
\fi
\begin{document}
%
\title{Towards a question answering assistant \\ for software development using a \\ transformer-based language model}

\author{\IEEEauthorblockN{Liliane do Nascimento Vale\IEEEauthorrefmark{1}\IEEEauthorrefmark{2}\\
Federal University of Catalão\IEEEauthorrefmark{1}\\
Institute of Biotechnology -
 Computing Science \\
Goiás - Brazil\\
Email: lilianevale@ufg.br}
\and
\IEEEauthorblockA{Marcelo de Almeida Maia}
Federal University of Uberlândia\IEEEauthorrefmark{2}\\
 Faculty of Computing\\
Minas Gerais - Brazil\\
Email: marcelo.maia@ufu.br}


%


\maketitle

\begin{abstract}
Question answering platforms, such as Stack Overflow, have impacted substantially how developers search for solutions for their programming problems. The crowd knowledge content available from such platforms has also been used to leverage software development tools. The recent advances on Natural Language Processing, specifically on more powerful language models, have demonstrated ability to enhance text understanding and generation. In this context, we aim at investigating the factors that can influence on the application of such models for understanding source code related data and produce more interactive and intelligent assistants for software development. In this preliminary study, we particularly investigate if a \textit{how-to} question filter and the level of context in the question may impact the results of a question answering transformer-based model. We suggest that fine-tuning models with corpus based on \textit{how-to} questions can impact positively in the model and more contextualized questions also induce more objective answers.
\end{abstract}


%
\IEEEpeerreviewmaketitle

\section{Introduction}
Question and answer (Q\&A) systems have gained expressive importance  over the past decade for software development. Stack Overflow (SO) is the most prominent example of such service, serving more than 100M users monthly, with more than 20M registered questions\footnote{https://stackoverflow.com/company}. The crowd knowledge available on Stack Overflow dumps have enabled several studies that leverage such raw content to produce documentation for APIs \cite{cookbooks2019}, recommend posts \cite{deSouza:2014:RCK, campos2016searching} and answers from queries \cite{silva2019recommending, crokage2020}, understand social interactions \cite{gender2019emse}, and several others\footnote{https://bit.ly/3oAbcUa}.

Recent advances in language models have opened the way for designing more intelligent Q\&A conversational assistants to perform tasks during the software development. Transformers, a special kind of neural network based solely on attention mechanisms, dispensed recurrence and convolutions and   have influenced a large number of other specific models \cite{attention2017nips}. Models  such as GPT-2 \cite{gpt2TR} and BERT \cite{bert2019} have demonstrated ability in several applications of Natural Language Processing (NLP), including Q\&A.

In this paper, we propose the investigation of how could an assistant based on transformers could perform  to answer automatically development related questions. We investigate the limitations on a model trained with the GPT-2\footnote{https://beta.openai.com/} to support the capacity to generate English text, fine-tuned with text (and code snippets) extracted from the SO dataset. Because question answering is still a hard challenge for English text in general, indeed, our goal is not to provide a functional assistant that could be readily used by developers, but instead to understand different factors that could impact on the accuracy of the answers, in order to guide further similar initiatives. 

In Section II, we describe the steps to train the assistant. In Section III, some preliminary results are presented and discussed. In Section IV, we present some related work, and finally in Section V, concluding remarks are presented.

\section{Study Setting}
Our proposal is to build a assistant that performs the task of automatically answering questions  using the GPT-2 language model fine-tuned with crowd knowledge. Thus, in the following sections we describe the steps necessary to conduct the proposed preliminary experiment. The data set used in this study is available in a Github repository\footnote{https://github.com/lascam-UFU/Towards-a-conversational-assistant-for-software-development-using-a-tr.git}.

\subsection{Goals}
The goal of this paper is not to propose a conversational assistant that could pass on a ``Turing test" customized for developers, i.e., developers would ask any question on a specific programming language or API, the assistant would answer, and the developer would not distinguish if it is a robot or a human. 
Our primary goal is to understand how a modern  transformer-based language model would perform in such a scenario, and more importantly, how different fine-tuning configurations and different questions may impact the result of the answers. So, our first goal is to understand if the fine-tuning with \textit{how-to}-oriented Q\&A pairs \cite{cookbooks2019} produces better answers for \textit{how-to} questions. Our second goal is to understand if providing more context in the question, i.e., a specific class or method, induces more contextualized answers.

\subsection{Corpus Preparation}
In order to fine-tune the GPT-2 model, we use questions and answers from SO. To build that corpus, initially approximately 20 million questions were recovered from the 2020 dump\cite{so}. For reasons of computational constraints, we select only the questions with the tag \#java and in English language. Java was chosen because of popularity and the number of systems developed with the language. Then, we selected 865k  Q\&A pairs with accepted answers ({\it AcceptedAnswerId}=1).
To structure the corpus for fine-tunning, we insert  the   $<|$startoftext$|>$,  $[$QUESTION$]$, $[$ANSWER$]$ and $<|$endoftext$|>$ delimiter tokens for each   Q\&A pair. These delimiters work as supervised training, allowing the model to learn the difference of question and answer, and consequently learn that the desired task is question answering. 

    
\subsection{Model Training}
Actually, the model training is  a fine-tuning of a pre-trained GPT-2 model. We use the pre-trained 124M model of the GPT-2 that makes it possible to learn the English language. We have chosen the smallest 124M model (default) because when fine-tuning GPT-2 because it balances speed, size, creativity and accuracy. Larger models would require more computational power. To fine-tune the model, we use  the corpus described in the previous section.
We have already built different text classifiers to distinguish the intent of a question in SO \cite{deSouza:2014:RCK, cookbooks2019}. A prevalent kind of intent the two set of questions is the \textit{how-to} questions, where the developers aims at solving a programming problem and expects a \textit{how-to} solution. So, we segmented the Q\&A pairs into two groups: one group (500k posts) containing questions with the word ''how'' on the posts named \textit{how-to} model, and one group (366k posts) not containing ``how'' on the posts named {\it non-how-to} model. Our goal is to compare the answers produced for the two fine-tuned models. 
The fine-tuning  loads the specified the SO corpus and trains for the specified number of steps. We maintained default of 1.000 steps, since it is enough to allow distinct text to emerge and takes just about 45 minutes. 
We used  Google Colab Notebook\footnote{https://colab.research.google.com/drive/1VLG8e7YSEwypxU-noRNhsv5dW4NfTGce?usp=sharing} with local connection in a particular server configured with 90Gb of RAM and 12 cores, and no GPU.

\subsection{Question Definition}
The fine-tuned models can be used to produce text given a preamble. In our case, the task is question answering, so the  preamble is a question and the generated text should be an answer for that question. In sequence, we established some criteria to define the questions for this study: 1)We use  ''how to'',  ''how can`` prefixes to compose questions; 2) We choose simple tasks on popular APIs/tasks, so we can more easily assess the quality of the answer produced by the model;
 3) To increase the  specificity of the context in the questions, we propose incremental questions: from a generic question, we added information progressively to evaluate how different could be the produced answers. So, we considered questions: 1) without context (in this case, we expect the model to suggest an API/class/method); 2) containing class name (in this case, we expect the model to explain the class and suggest methods of that class); 3) with a class and a method name (in this case, we expect the model to explain how to use that specific method). The following questions considered in the study are shown:\\
{\scriptsize  \texttt{
Q1: How to create a window?\\
Q2: How to create a window with JFrame \\
Q3:How to create a window with JFrame and set dimension? \\
Q4:How to open a DataBase connection? \\
Q5:How to open a DataBase connection JDBC?\\
Q6:How to open a DataBase connection using JDBC and DriverManager?\\
Q7: How to open a DataBase connection using JDBC and DriverManager and getConnection method?\\
Q8: How can I insert an element in array at a given position?\\
Q9: How to open a SSL connection?}}

\subsection{Answer Generation}
For the generation of responses by the GTP-2 model, we maintained the prefix to specify how exactly the text must start (example: [QUESTION]: How to open database connection? $\verb!\n!$
[ANSWER]:"), in this case by the question itself. 
 The answers produced had a length equals to 250 tokens and to generate the amount of text generated: size equals to 5; number of samples equals to 5 and temperature equals to 0.7. An increase of  temperature would increase creativity of the text, however we would prefer less creativity trying to avoid meaningless solutions.

\section{Results and Discussion}
In all provided answers, we could not find a completely right one. Next, we report on some encouraging answers, together with interesting failures in the answers.
Initially, we will consider the following the Q1, Q2, and Q3 from the previous section.
Regarding the \textit{how-to} model,  answers for Q1 does not mention specific classes for instantiating windows as JFrame, JPanel. The best that was generated was a suggestion involving  \textit{paintComponent (Graphics g)}.

{\scriptsize
\begin{verbatim}
public class Window {      
    private String userName; private String password;      
    public Window(String userName, String password) {              
        this.userName = userName;     }     
    @Override public void paintComponent(Graphics g) {        
         super.paintComponent(g);     } 
\end{verbatim}
}

Answers, for both Q2 and Q3, present examples of instantiation and use of the JFrame class, but for Q3 there are no specific methods for determining the window size, such as the \textit{setPreferredSize()} or \textit{setSize()} methods. We show part of the first answer for Q2, which mostly seems to make sense, and also has learned adequate syntax, although with some errors. This is a promising example:
{\scriptsize
\begin{verbatim}
JFrame mainFrame = new JFrame("Main");            
mainFrame.setDefaultCloseOperation (WindowConstants.EXIT_ON_CLOSE); 
mainFrame.setLayout(new BorderLayout());         
JFrame mainFrame.add(mainFrame);         
mainFrame.add(new JLabel("Main"));         
\end{verbatim}
}

Regarding the \textit{non-how-to} model, only Q2 has an answer that refers to the \textit{JFrame} class and the \textit{setSize()} method. For Q3, there is a reference to the \textit{JPanel} class. For the generic Q1, there is a mention of the use a non-existent $javax.swing.WindowBuilder$ class for building windows. This is an interesting limitation. If in English, the use of new entities would be acceptable in some contexts, in software development the use of new entities seems to require the proper definition of that entity.

For the second group of questions related to database connection (Q4, Q5, Q6, and Q7), we have the following findings for the \textit{how-to} model.

Q4 did not indicate an API to use. The suggested code made a confusion of connection with databases and network connection:

{\scriptsize
\begin{verbatim}
 client.connect(WebSocket.getConnectionContext());     
 ConnectionFactory factory = new ConnectionFactory();              
 connectionFactory.setConnection(new WebSocketConnection(client)); 
\end{verbatim}
}

Q5 specifies that JDBC should be used, and presented examples of instantiating  \textit{Connection}, which is not a class but an interface. Moreover, it proposed a snippet to setup users and password, which are entities related to a connection.
Q6 mentioned the class \textit{DriverManager} and showed better results containing code examples involving \textit{Connection} and  \textit{DriverManager} classes and \textit{getConnection()} method and SQL:

{\scriptsize
\begin{verbatim}
DriverManager.getConnection().reset(); 
 public class DatabaseTableAdapter extends JTableAdapter { 
 private static final String SQL_NAME_TYPE = "sql"; 
\end{verbatim}
}

 Q7 presents examples of use of the $DriverManager$ class, but all inadequate. The snippet use a non-existent $setConnection()$ method, suggesting the existence of the $getConnection()$ method. Moreover, \textit{DriverManager} was not used properly accessed based on its static methods.
 
{\scriptsize
\begin{verbatim}
  DriverManager implements DriverManager {       
  @Override public void onDriverManagerChanged 
  (DriverManager driver) { driver.setConnection
   (driver.getDriverManager());   driver.setDataSource(dataSource);}
   @Override public void onDataSourceChanged (DataSource source) {       
      connection = driver.getDataSource();  
 \end{verbatim}
}
 
Regarding \textit{non-how-to} model, Q4, which is more general question, suggests the use of MySQL and Q5 refers to a webservice.
  Q6 cites the class \textit{DriverManager} but with improper and meaningless methods:
  
{\scriptsize
\begin{verbatim}
  String path = DriverManager.getSerializedPath(); 
  String uri = driverManager.getSerializedURI();     
  String key = driverManager.getSerializedKey()
\end{verbatim}
} 

Q7, which puts \textit{DriverManager} and \textit{getConnection} in context, also shows the invocation of some methods, but also improperly, although with more sense than in Q6:

{\scriptsize
\begin{verbatim}
  qlConnection.ConnectionBuilder());
  driverManager.setConnection(driverManager.getConnection()); 
  driverManager.setConnection (driverManager.getConnection()); 
  DriverManager driverManager = DriverManager.getInstance();  
  DriverManager.createConnection();
\end{verbatim}
}

 In Q8, regarding the insertion of elements in an array: {\it How can I insert an element in array at a given position? }, the \textit{how-to} model retrieved an example related to arrays, but still not a proper one.

{\scriptsize
\begin{verbatim}
for (int i = 0; i  num.length; i++) { 
  for (int j = 0; j  num.length; j++) {       
  if (num[i] & 0) {  if (num[j] & 0) {             
        int num[num.length - j] =   num[j].length  }    }          
      for (int k = 0; k  num.length; k++) {        
        if (num[k] & 0) {              
           int num[k + j] = num[k + j + 1]; }       
\end{verbatim}
}
 
Regarding to the  \textit{non-how-to} model, we  observed no satisfactory answers on this group.
 Q9 is a generic question (\textit{How to open an SSL connection?}), and both models did not presented adequate answer, but a possible answer that can be a guide to formulate new queries  was founded in the ``how-to" model: ''\textit{Open a ssl connection to your hosting provider.  The following steps will help you:   Open a ssl connection to the URL of the ssl project. In the ssl connection, add the URL of the ssl project to the URL of the ssl server}.`` 
 For the \textit{non-how-to}  model,  the best answer was: ``\textit{The Java API is to be used in the context where the client is executing.  In this case, you need to use a client-side method to send the request to the server.     So basically, you don't need to do any of this, because it's just a wrapper around the client.      You should have a single call to send the request to the server.      If you want to use a client-side method, then you should use a client-side method to send the request to the server.  Note the static method that returns the client-side method. ...}''. Although related to the network topic, it lacks meaningfulness.
    
We used two different models to evaluate the generated answers, and then compare their performance. We could observe that the \textit{how-to} model, compared to the \textit{non-how-to},  was able to produce more contextualized answers, sometimes containing legible pieces of code and consistent explanation. 
One possible reason for that better performance may related to the structure of the question. We selected posts that could contain the word ''how`` in any part of the text. Actually, we observed that most posts started with the word ''how``. Thus, it is possible that this structure allowed the model to better relate the intent of the questions, inducing their semantic understanding.

Regarding to our second goal, we observed that the generation process had difficulties in answering generic questions, even though the posts are specific to the Java language. Therefore, we understand that query reformulation methods would be important to recommend more specific questions allowing the assistant to be more accurate in the answers.

A point observed was the quality of the answers in terms of grammar, semantic coherence, etc. We understand that  the pre-trained 124M model for English text works quite well for producing apparently coherent text, but still there is much to be done to achieve more semantic accuracy.

\section{Related Work}
IntelliCode \cite{SvyatkovskiyDFS20} it is  a general purpose multilingual code completion tool which is capable of predicting sequences of code tokens of arbitrary types, generating up to entire lines of syntactically correct code. It leverages generative transformer model trained on 1.2 billion lines of source code in Python, C\#, JavaScript and TypeScript programming languages.
Other commercial tool, Tabnine\footnote{http://www.tabnine.com} uses GPT-2 to serve ranked lists of code sequence suggestions. However,this tool does not attempt to complete longer sequences of 20–30 characters long, up to a whole line of code, and we are not aware of any currently deployed tool that have similar feature.

 \citet{core}, a novel deep learning model for recommending relevant reviews given a code change, named as COde REview engine - CORE that is built upon only code changes and reviews without external resources.
The motivation is to reduce the workload of developers by providing review recommendation without human intervene. By automating the code review process, developers can correct their code as soon as possible, hence, reducing the time between each revision of code changes.

\citet{gpt2TR}  demonstrate that language models begin to learn these tasks without any explicit supervision when trained on a new data set of millions of webpages called WebText. When conditioned on a document plus questions, the answers generated by the language model.
 Their idea was to make a model unsupervised "independent task", but a difficult problem to apply this unsupervised "task independent" model in a practical solution. To improve the results are gave some clues of the task for the model: 1)  summary was added the token TL; DR: and one more attempt of summary in the training; 2) in translation the authors pass the task in the context 3) Q$\&$A induce the task also in the context that authors passed.
 
The most of the approaches that employ language and learning models have been used mainly to recover code. Our proposal presents an alternative to recover code and text that helps to explain and learn a given concept.

\section{Conclusion}
We have presented a preliminary study aimed at investigating the use of a transformer-based language model aiming the development of  a question answering assistant for problems  related to software development. 
Our findings suggest that the corpus used in fine-tuning should be carefully designed, so the answers can be more accurate. Moreover, as expected, we found that the question formulation still plays a very important role on the generated answer. For future work, several others factors still need investigation. For instance, a lower level of creativity may  influence the generation of more consolidated solutions. Other language models, including newer generation GPT-3 trained with more parameters, should be investigated on their impact on the quality of the generation.


\section*{Acknowledgment}
The authors would like to thank CNPq and FAPEMIG for partially supporting this work.



\bibliographystyle{IEEEtranN}
\bibliography{biblio}
%



\end{document}